\documentclass[]{spie} 
\pdfoutput=1 
\usepackage[]{graphicx}
\usepackage{subcaption}
\usepackage{xcolor}

\usepackage[utf8]{inputenc}
\usepackage[T1]{fontenc}
\usepackage{parskip}
\usepackage{wasysym}
\usepackage{multirow}

\usepackage{graphicx}
\usepackage{media9}
\usepackage{cite}
\usepackage{hyperref}

\title{Performance of Large-Format Deformable Mirrors Constructed with Hybrid Variable Reluctance Actuators II: Initial Lab Results from FLASH}

\author{Rachel Bowens-Rubin\supit{a}, Daren Dillon\supit{a}, Philip M. Hinz\supit{a}, 
Stefan Kuiper\supit{b}
\skiplinehalf
\supit{a}UC Santa Cruz, 1156 High St, Santa Cruz CA, USA; \\
\supit{b} TNO Technical Sciences, Delft, The Netherlands
}
\authorinfo{Further author information: (Send correspondence to R. Bowens-Rubin)\\R.B.R.: E-mail: rbowru@ucsc.edu}

\begin{document} 
\maketitle 
  
  \begin{abstract}
 Advancements in high-efficiency hybrid variable reluctance (HVR) actuators are an enabling technology for building the next generation of large-format deformable mirrors, including adaptive secondary mirrors. The Netherlands Organization for Applied Scientific Research (TNO) has developed a new style of hybrid variable reluctance actuator that requires approximately seventy-five times less power to operate as compared to the traditional style of voice-coil actuators. We present the initial performance results from laboratory testing of TNO's latest 19-actuator prototype deformable mirror, FLASH. We report the actuator cross-coupling,  linearity, hysteresis, natural shape flattening, and drift as measured with a Zygo interferometer and a set of four capacitive sensors.  We also present results of the dynamic performance of the FLASH on sub-millisecond timescales to estimate the limits of this technology for high-contrast imaging adaptive optics. We confirm that this technology has strong potential for use in on-sky adaptive secondary mirrors without the need for active cooling. 

\end{abstract}

 
\section{INTRODUCTION}

\indent \indent Adaptive secondary mirrors have been used successfully at telescopes like the Large Binocular Telescope~\cite{Esposito2010}, Magellan~\cite{Close2018}, the VLT Observatory~\cite{Biasi2012}, and the MMT Observatory~\cite{Wildi2003}. However, these adaptive secondary mirror systems have historically been complex to build and operate.  The listed adaptive secondary mirrors are constructed using power intensive voice-coil actuators that require a dedicated cooling system to dissipate heat.  They also require closed-loop capacitive sensors to monitor the mirror position and correct for nonlinearity and hysteresis in the actuator displacements.~\cite{Hinz2016} The mirror facesheet fabrication has  traditionally required the grinding of a large piece of glass into a thin glass sheet that is flexible enough to be shaped by the actuators. 

The Netherlands Organization for Applied Scientific Research (TNO) has made significant breakthroughs in large-format deformable mirror technology that could enable adaptive secondary mirrors to become simpler and less costly to operate.   The key advancement is a new style of hybrid variable reluctance (HVR) actuator that is $\sim$75 times more efficient than traditional voice-coil actuators (HVR actuator efficiency~\cite{Kuiper2018} $= 38 N/\sqrt{W}$; MMT and LBT actuator efficiency~\cite{Riccardi2003} $= 0.5 N/\sqrt{W}$). The low required power offers a pathway to eliminate the actuator cooling system, miniaturize the system, and pack the actuators at higher density. TNO is also developing solutions, in conjunction with partners and the Fraunhofer Institute and the University of California Santa Cruz, that could simplify the facesheet fabrication process through hot-forming thin, commercially-available glass sheets. 

Adaptive secondary mirrors using the HVR technology are being proposed for use in three telescopes: the University of Hawaii 2.2-meter (UH2.2m), the University of California Observatories Automated Planet Finder Telescope (APF), and the W.M. Keck Observatory (Keck). 
The UH 2.2m ASM will be the first on-sky operation of the TNO technology and is expected to be commissioned during 2022~\cite{Chun2020}. ASM's for the APF and Keck are in the proposal phase.  More information regarding the science case and initial design concepts for the APF ASM can be found in Hinz et al 2020~\cite{Hinz2020}.   Additional information regarding the Keck ASM can be found in the Keck White paper, \emph{An adaptive secondary mirror concept for the W.M. Keck Observatory based on HVR technology} by Hinz et al.

While the TNO large-format mirrors show great potential to make advanced adaptive optics systems accessible to a wide variety of observatories, they have yet to be proven to the astronomical community. 
This paper presents the performance testing  of TNO's fourth deformable-mirror prototype, the First Lab for adaptive optics Adaptive Secondary Holophote (FLASH). 
It is the second SPIE paper in a series, following Bowens-Rubin et al. 2020 which reported the performance testing results from TNO's third large-format deformable mirror (DM3).~\cite{BowensRubin2020}

\section{TECHNOLOGY OVERVIEW}

\subsection{19-Actuator FLASH Prototype}

\begin{figure} [h!]
  \centering
  {\includegraphics[width=0.5\textwidth]{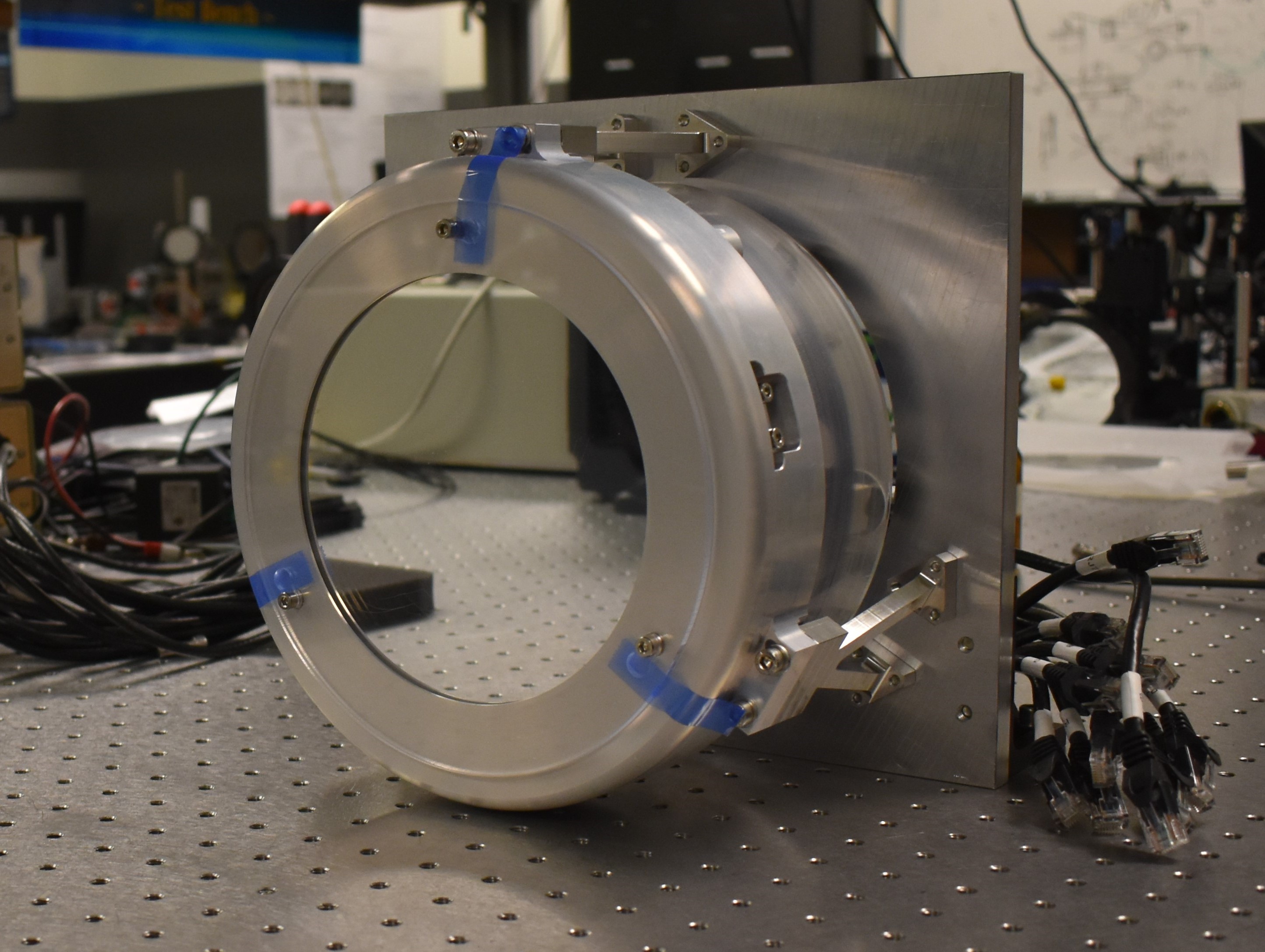}}~
  {\includegraphics[width=0.458\textwidth]{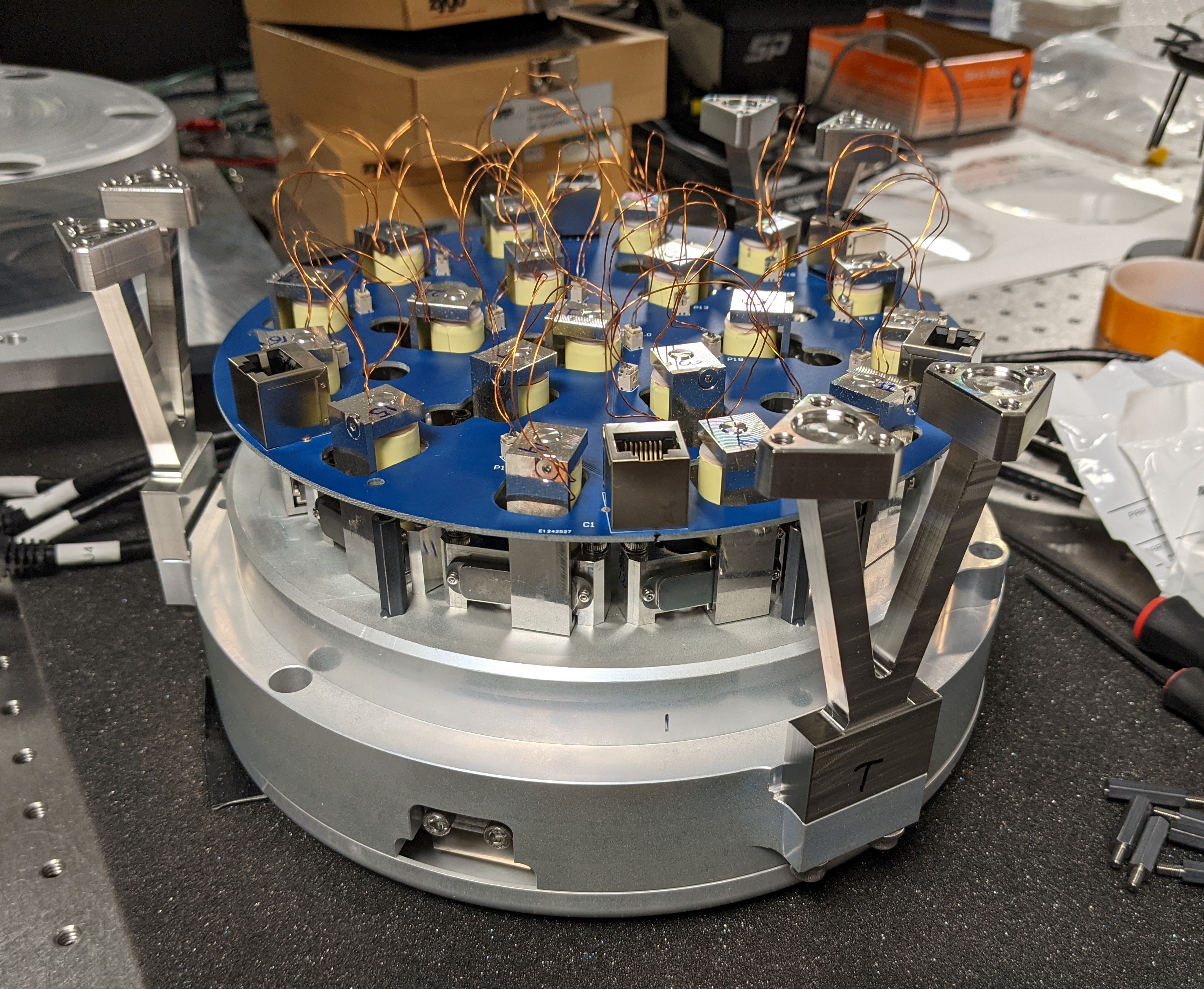}}~
 \caption{ \textbf{FLASH 19-Actuator Large-Format Deformable Mirror}. (\textit{Left}) The FLASH mounted on the testbench at the UC Santa Cruz Lab for Adaptive Optics. It is pictured  here with its protective mirror-facesheet cover on. (\textit{Right}) Back side of the FLASH mirror with the actuators and electronics breadboard. }
     \label{fig:FLASH}
\end{figure}

\begin{figure}[h!]
   \begin{center}
   \begin{tabular}{c}
   \includegraphics[width=0.8\textwidth]{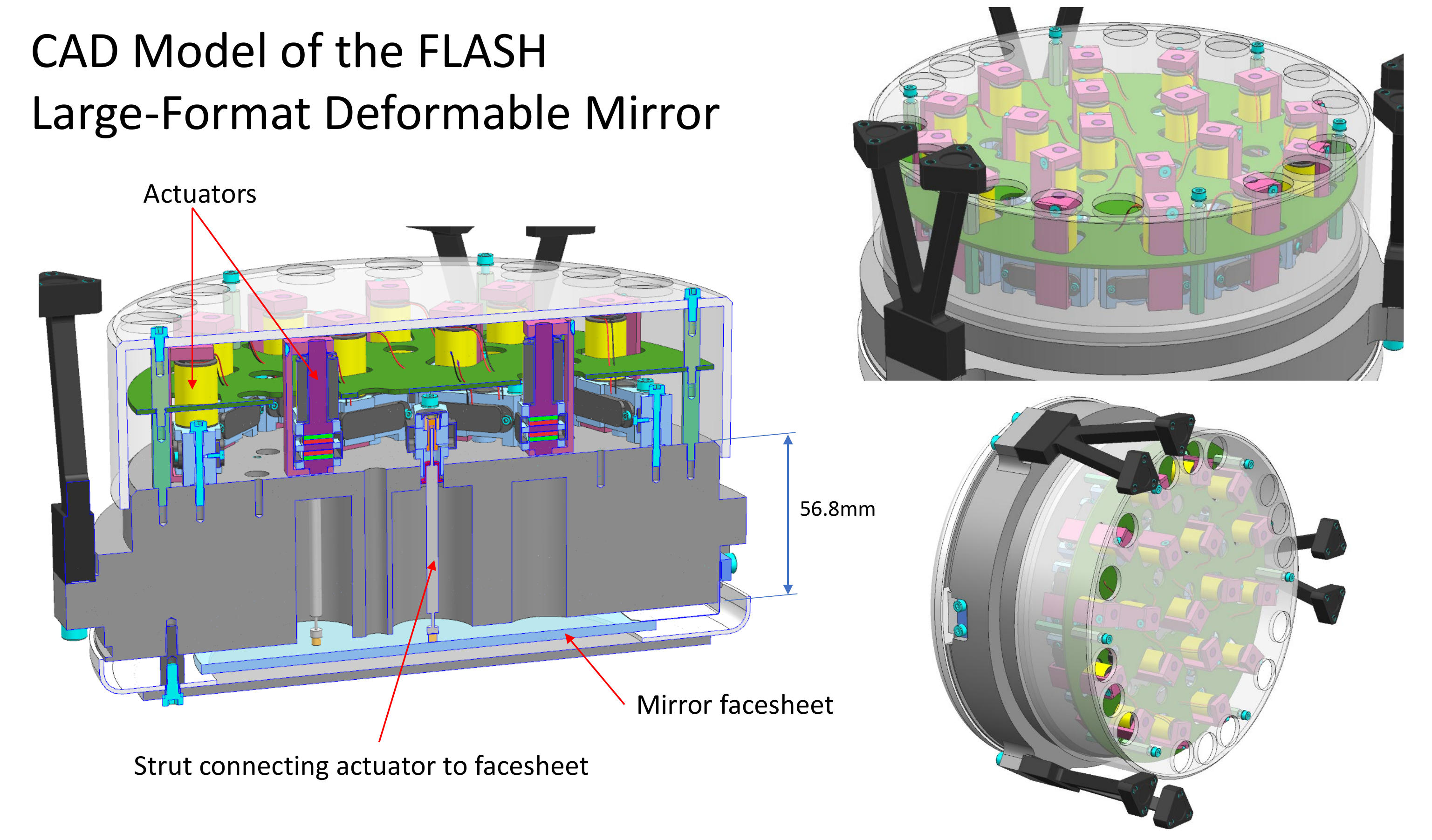}
   \end{tabular}
   \end{center}
   \caption[example] 
   { \label{fig:FLASHCAD} 
\textbf{CAD Model of the FLASH Large-Format Deformable Mirror}.  FLASH is designed with an aluminium backing structure which supports the mirror facesheet, the actuators, and the electronics breadboard.  }
   \end{figure}

The FLASH is the fourth prototype large-format deformable mirror built by TNO (Figure \ref{fig:FLASH}). The FLASH was commissioned by the University of California Lab for Adaptive Optics (UCSC-LAO) to verify the performance of the current generation of TNO technology before the construction of the UH2.2m ASM.  It was constructed in 2020 and delivered to the UCSC-LAO in January of 2021. 
The mirror is a companion to the UH 2.2m adaptive secondary mirror~\cite{Chun2020} and  shares the same facesheet thickness (3.3mm), actuator spacing (39mm), and layout geometry (hexapolar).   The UH2.2m ASM and FLASH share the same generation of TNO HVR actuator, which is described in more detail Section 2.2. 
Figure \ref{fig:FLASHCAD} shows the internal designs of the FLASH, and Table 1 summarizes the FLASH design specifications.  

FLASH is controlled through a 64 analog output from a Real Time Linux environment. The user can control the shape of the mirror's surface by specifying the value of the current applied to each of the 19 actuators in an RDA interface in Matlab running on an office PC.  The actuators are plugged into a breadboard. The breadboard and control electronics are connected by ethernet cables. Each ethernet cable controls the signal to four actuators.   The electronics can produce an applied voltage of up to +10V across each actuator.
\footnote{There is a correction for Bowens-Rubin et al. 2020 related to the voltage applied across each actuator.  It was discovered in 2021 that a driver setting had previously limited the electronics to producing +5V during the DM3 testing instead of +10V. This caused all currents reported for the DM3 to be stated as doubled their actual value.  If comparing the results from the DM3 in Bowens-Rubin et al. 2020 to the FLASH, all currents reported for the DM3 should be halved.   }

\begin{center}
\begin{table}[h!]
\caption{\textbf{FLASH Specifications}}
\centering
\begin{tabular}{|l|c|}
\hline
Total Diameter                       & 160mm     \\ \hline
Facesheet Diameter                   & 150mm     \\ \hline
Number of Actuators                  & 19        \\ \hline
Actuator Spacing                     & 39mm      \\ \hline
Actuator Generation (year)           & 2020      \\ \hline
Actuator Layout Geometry             & Hexapolar \\ \hline
Facesheet Thickness                  & 3.3mm     \\ \hline
Facesheet Material                   & Borofloat \\ \hline
Control Electronics Used for Testing & Analog    \\ \hline
Available Capacitive Sensor Locations       & 13        \\ \hline
\end{tabular}
\end{table}
\end{center}

\subsection{2020 Design of the TNO Hybrid Variable Reluctance Actuator}

\begin{figure} [h!]
  \centering
  {\includegraphics[width=0.40\textwidth]{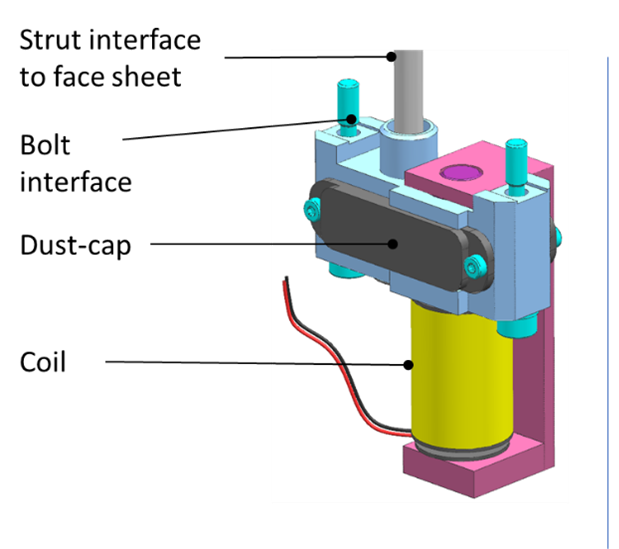}}~                
  {\includegraphics[width=0.30\textwidth]{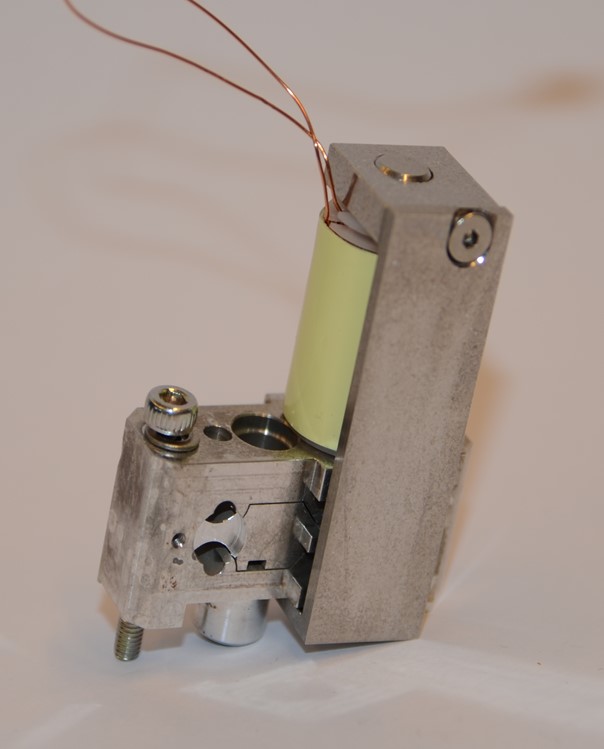}}~
 \caption{\textbf{TNO's 2020 version of the HVR Actuator}  (\textit{Left}) CAD model of the TNO HVR Actuator designed for FLASH and the UH2.2m adaptive secondary mirror. (\textit{Right}) A picture of the realized FLASH actuator from the  batch manufactured in November 2020. }
     \label{fig:2020act}
\end{figure}

The FLASH is constructed with 19 hybrid-variable reluctance (HVR) actuators (Figure \ref{fig:2020act}). The HVR actuators were designed by TNO and constructed by VDL-ETG in the Netherlands. The 2020 actuator design was based on the actuators developed for TNO’s  57-actuator DM prototype (DM3)~\cite{Kuiper2018}.  While previous generations of the HVR actuators were made as multiple units in a strip, the 2020 actuators are individual freestanding units that can be bolted on to a supporting body using two bolts at any spacing greater than the actuator's diameter. Dust caps were added to the 2020 generation to prevent dust from reaching vital parts inside the actuators. Provisions were also made in the actuator design to allow quick replacement in the event of an actuator failure. The actuator replacement procedure was tested during the construction of the FLASH, and it was completed in under four hours.

The force output of the 2020 HVR actuators is $\pm$8N over a linear range (99.5\%) before entering in magnetic saturation and can go up to a maximum force of $\pm$14N when fully magnetically saturated.  Modifications from the previous generations were made to the internal mechanical hinges and the internal lever-arm to match the different dynamic characteristics of the deformable mirror system. This internal lever-arm transmits the force produced by the actuator to the strut that interfaces with the facesheet with an amplification factor of 2.6, which  results in a force linear range of $\pm$21N on the facesheet. The modeled  free displacement range of the actuator is 35-40µm (PV), and the modeled  inter-actuator stroke is $\sim$5µm (PV). The average resistance measured for each actuator was $R = 3.68\pm0.07\Omega$, and the average inductance measured for each actuator was $I = 14.85 \pm 0.07$mH.

\subsection{Facesheet Glass-Slumping Experiments} 

The facesheet for FLASH was constructed from a 3.3mm-thickness, 160mm-diameter borofloat wafer supplied by Swift Glass of Elmira, NY.  A set of ten wafers were supplied and tested for flatness using a Mercury vapor illumination screen and a $\lambda$/10 reference flat.  The wafers all showed 
5.5-9.5$\mu$m
of wavefront error, which appeared as convex curvature on one side. A similar amount of concave curvature with edge rolloff was visible on the other side.  The UCSC-LAO attempted to flatten these wafers further via heating in a glass kiln. This is part of a larger program to eventually produce curved facesheets for large format deformable mirrors using initially flat borofloat sheets and a hot forming process, similar to approaches developed for X-ray mirrors (REF's).  

These initial tests used a fused silica reference flat (the same one used for measuring flatness) as the bottom mould for the facesheet.  Improved figure was attempted by various tests that used the self weight of the shell to slump, as well as additional fused silica top weights.  While reduction of the curvature was achieved, the resulting facesheets had several waves of error on spatial scales of 20-30mm remaining. This was judged to be more problematic for FLASH operation than the initial larger, but smoother wavefront error on the unslumped shells.  Therefore the final facesheet selected for integration was one that was not heated and was shown to introduce $2.8\mu$m peak-to-valley convex curvature of surface error.

The experience with hot forming a facesheet for FLASH lead us to several conclusions for future slumping work.  Primarily, it was found that there is a very narrow range of temperatures for which the shell becomes soft enough to take on the shape of the mould but is cool enough to not stick and destroy the mould and the facesheet.  A release agent (such as boron nitride) might help with this but may lead to high spatial-frequency structure being imprinted on the glass.  Top loading of the shell can help when using cooler temperatures, but then the accuracy of the surface for the top weight becomes important.   Further, cleanliness of the bottom mould is critical to avoid localized bumps from any particles caught between the mould and the slumped facesheet.  Since heating is typically carried out in a glass kiln that uses ceramic bricks for insulation, a clean environment will require a more complicated setup.   As we scale up such an approach, a significant portion of the cost becomes making the precision mould. 

The above process might be simplified by removing the mould and letting the facesheet slump freely, using a ring slightly smaller than the diameter of the facesheet for bottom support. Using no top weight results in a curve with higher order radial terms once the glass is heated. A suitably placed ring top weight can provide loading on the facesheet, such that the final curve approximates the desired conic shape to within the precision needed for the actuators to remove the residual error.  The UCSC-LAO is exploring this approach for future large-format curved facesheets.  

The FLASH facesheet was aluminium coated on the front and back surfaces.  The back was coated to minimize the optical warping due to having a mixed material surface. The back coating also provides a conductive surface for the capacitive sensors to measure against. The areas on the back of the facesheet that would be bonded to the actuator struts were not coated.  Circles of $\diameter$9.5mm kapton tape were adhered before coating to prevent these areas from being aluminized. 

\subsection{Zygo Interferometer Measurement Setup}

\begin{figure}[h!]
   \begin{center}
   \begin{tabular}{c}
   \includegraphics[height=7cm]{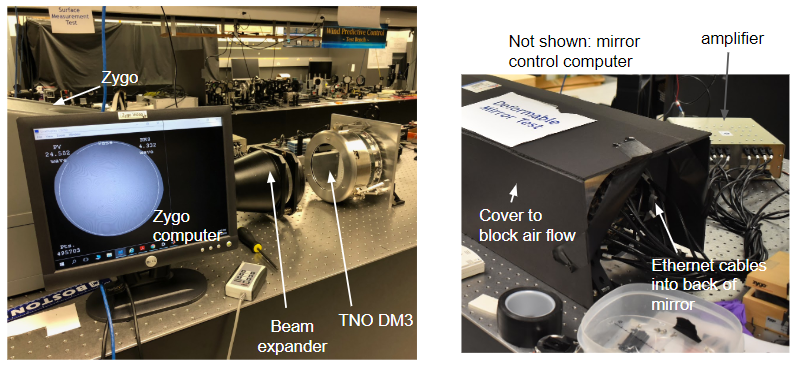}
   \end{tabular}
   \end{center}
   \caption[example] 
   { \label{fig:testbed} 
\textbf{Large-Format Deformable Mirror Testbench at the UCSC Lab for Adaptive Optics}. The testbench utilizes a Zygo Interferometer with MetroPro software to measure the surface shape. A beam expander is used so that the majority of the surface of the FLASH can be measured in a single image (up to $\diameter$140mm).  }
   \end{figure}

 The large-format deformable mirror testbed at UCSC-LAO utilizes a Zygo interferometer that is controlled using Metropro Software (Figure \ref{fig:testbed}). It can measure the shape of the mirror’s surface to a precision of 0.6nm of surface error within $\diameter$140mm. Measurements can be collected at a maximum speed of one image per second. 
 
 The Zygo interferometer is not capable of measuring the piston changes in an optical system.  To calibrate for  piston, we measure the median value of the image and subtract it. We filter the empty pixels (listed in the Zygo data as 'NaN' values) by intrapolating the values using nearby pixels.   We convert the measurements reported by the Zygo interferometer in units of `Zygos' to units of  nanometers of surface error by
 
 \begin{equation}
     PhaseData[nm] = PhaseData[zygos] \frac{\lambda S}{R}
 \end{equation}
 
\noindent where the wavelength of the laser in the Zygo interferometer is $\lambda = 632.8$nm, the geometry factor is $S = 0.5$, and the resolution setting (high) is $R = 32768$ (Zygo interferometer manual: zygo-0347M, 12-5 p447).  We analyze the Zygo interferometer data using Matlab and load the images using the \textit{LoadZygoBinary} function authored by Massimo Galimberti.

\subsection{Capacitive Sensor Measurement Setup}

FLASH is the first large-format deformable mirror from TNO to incorporate space for an internal capacitive  sensor system.  While the Zygo interferometer measurement setup can be used to gain an understanding of the full shape of the mirror's surface, it cannot collect time series data.  
The capacitive sensor system can provide data with a readout speed of 3906Hz over a working range of 50$\mu$m.  This fast frame rate is needed for evaluating if this technology is suitable for use in fast adaptive optics correction and high-contrast imaging.  However, the capacitive sensor data are  limited spatially, capturing displacement information only in the small coverage area that they are located. To determine if having limited access to a few points on the surface is an adequate proxy for the full mirror behavior, our linearity, hysteresis, and drift testing were performed with both the Zygo interferometer measurement setup and capacitive sensor systems side-by-side. 
If successful, future testing when an interferometer is prohibitively difficult to use may rely solely on the capacitive sensor system to understand the behavior of the system.    

\begin{figure}[h!]
 \centering
  {\includegraphics[width=0.42\textwidth]{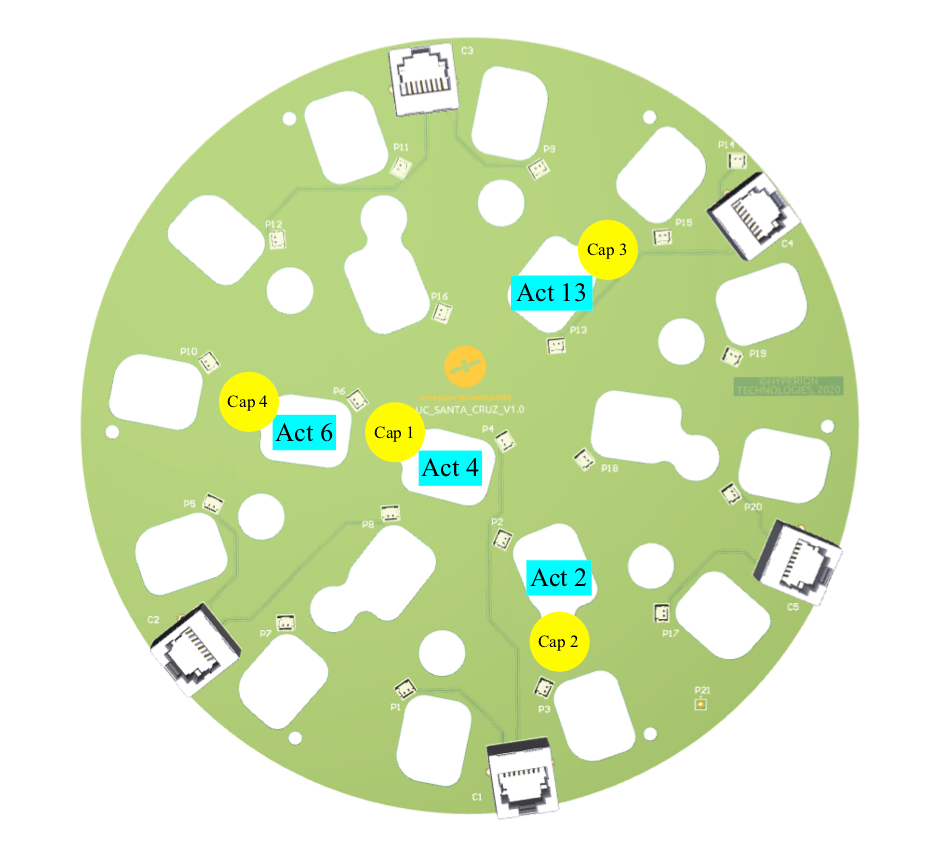}}~        
  {\includegraphics[width=0.56\textwidth]{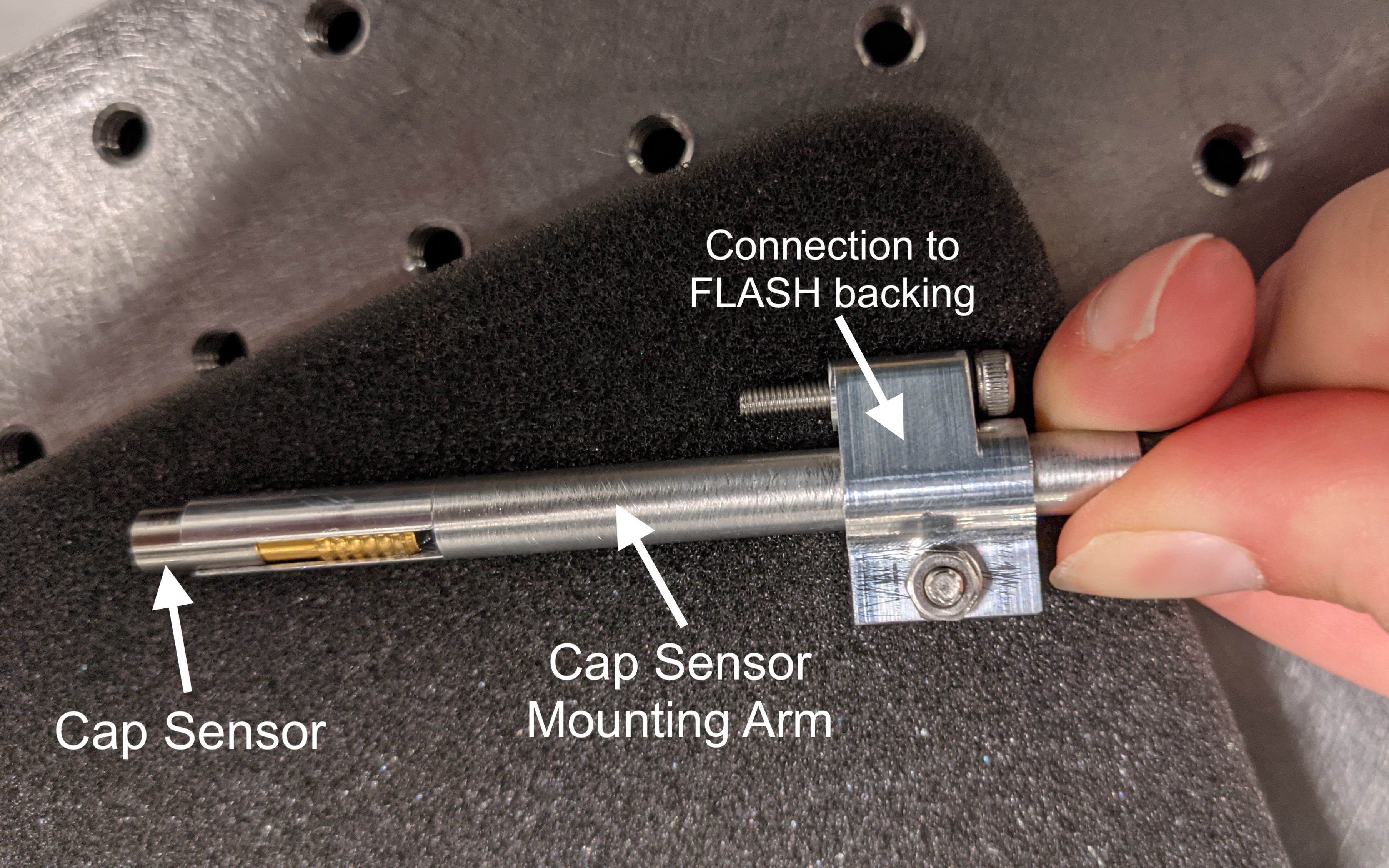}}~
   \caption { \label{fig:capsensor}
\textbf{FLASH Capacitive Sensors}. (\textit{Left}) Four capacitive sensors were added to the internal structure of FLASH for testing at UCSC-LAO.  The sensors were placed next to the actuators that were used in linearity and hysteresis testing. (\textit{Right})  A capacitive sensor inside its mount before it was installed.  } 
   \end{figure}

The FLASH backing structure has thirteen holes available for capacitive sensors to be mounted. 
Four Micro Epsilon CS005 capacitive sensors were internally mounted for testing the FLASH at the UCSC-LAO (Figure \ref{fig:capsensor}, \textit{left}). The sensors were run using the Micro-epsilon capaNCDT 6200 controller and the Micro-epsilon web interface software in an open loop system.

The mounts to position the capacitive sensors directly behind the mirror facesheet were designed by UC Santa Cruz. They were manufactured out of aluminum and are are 60mm in length.  The capacitive sensor is glued in place on the mounting tube (Figure \ref{fig:capsensor}, \textit{right}).  To install the capacitive sensor mounts in the FLASH backing structure, the actuators were unplugged from the breadboard and breadboard was removed and reinstalled.  The capacitive sensor mount installation was completed in approximately three hours.

\newpage

\section{FLASH PERFORMANCE TESTING RESULTS}

\subsection{Influence Function Profile and Actuator Cross-Coupling}

\emph{Measurement setup: Zygo Interferometer }

\begin{figure}[ht]
   \begin{center}
   \begin{tabular}{c}
   \includegraphics[width=0.99\textwidth]{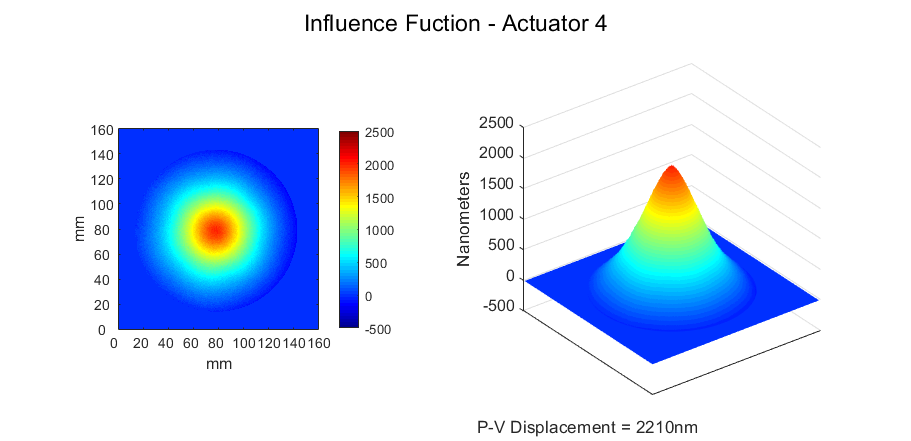}
   \end{tabular}
   \end{center}
   \caption[example] 
   { \label{fig:exampleinflufun} 
   \textbf{Example Influence Function}: An applied voltage of +50mA to Actuator 4 resulted in an actuator displacement of 2210nm.  These images are scaled with the x-y axis in millimeters and the color axis and z-displacement in nanometers.  The natural shape of the FLASH mirror was subtracted, so only the displacement due to the actuator poke is visible.  }
   \end{figure}

The influence function of each actuator was measured using an individual actuator poke test.  A $+50$mA current was applied in an automated measurement sequence to displace each of the 19 actuators one-by-one. The displacement of these pokes corresponded to approximately 2200nm peak-to-valley.   Eighteen actuators responded to the positive current with a positive displacement, and one actuator responded with a negative displacement. Seven of the actuators had profiles that were fully visible in the field of view of our Zygo interferometer setup: the center actuator (Actuator 4) and the six actuators in the middle ring (Actuators 2, 6, 8, 13, 15, and 17). 
An example of the influence function measured for Actuator 4 is shown in Figure \ref{fig:exampleinflufun}.  

\begin{figure}[h!]
 \centering
  {\includegraphics[width=0.40\textwidth]{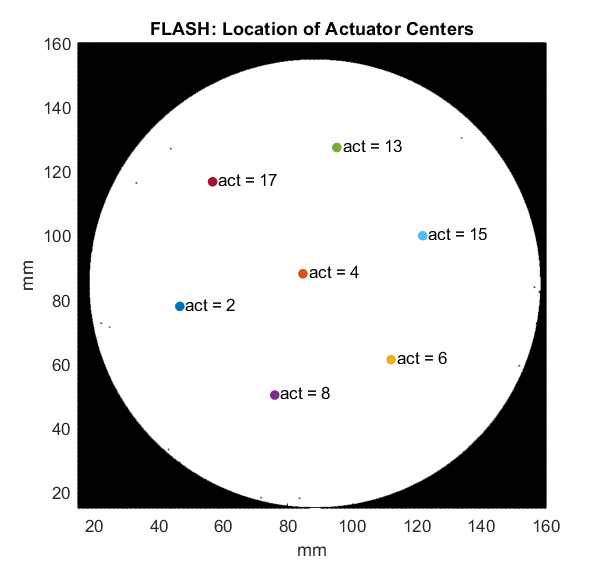}}~    
  {\includegraphics[width=0.60\textwidth]{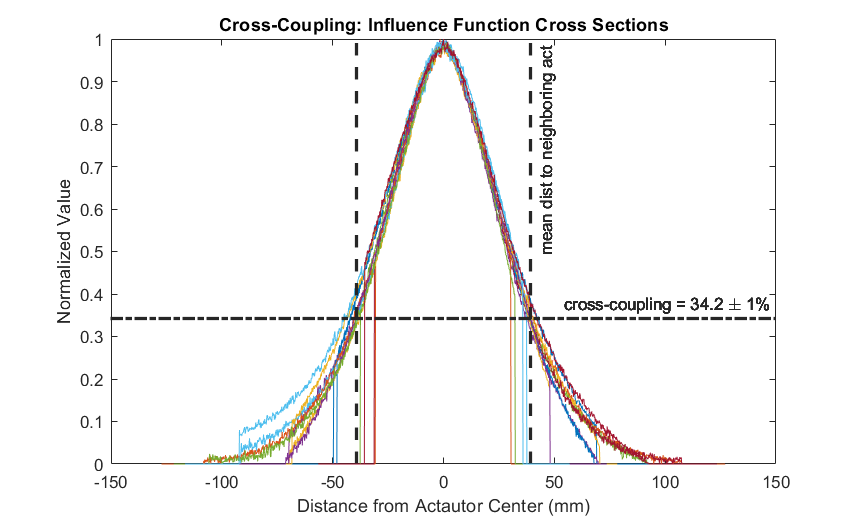}}~
   \caption { \label{fig:crosscoupling} 
\textbf{Actuator Cross-Coupling}. (\textit{Left}) Seven actuators were used to calculate the cross-coupling between actuators.  The centers of these actuators were fully visible within the frame of the Zygo interferometer.  (\textit{Right})  To measure the cross-coupling the actuator, 14 cross sections from the column and row direction from each visible actuator were aligned at their centers and normalized.  The average distance between the middle ring actuators to the center was 39.4 $\pm$ 0.4mm. This corresponds to a cross-coupling measurement of 34.2 $\pm$ 1\%. } 
   \end{figure}

The actuator cross-coupling was measured to be 34.2$\pm1\%$.  This value was determined using the influence functions from the seven actuators that were fully visible in the Zygo field of view (Figure \ref{fig:crosscoupling}).  The location of the center of each of the seven actuators in the image was found using a Gaussian fit. A cross-section of the influence function was then taken in the row and column directions, emanating from the actuator center location. The fourteen cross sections were then normalized and stacked.  The average distance between the center actuator to each actuator in the inner ring was then determined using the location of the centers of the actuators and measured to be 39.4 $\pm$ 0.4mm. The value of the fourteen normalized cross-sections at the width location of average actuator separation were averaged to yield the cross-coupling measurement.  This measurement is consistent with the cross-coupling value expected from the pre-fabrication modeling done by TNO.

\subsection{Natural-Shape Surface Flattening} 

\emph{Measurement setup: Zygo Interferometer }

The natural shape of the FLASH mirror facesheet has a peak-to-valley that is 5339.1nm with a 1158.8nm RMS variation (within $\diameter$140mm). To bring the surface as flat as possible (as measured by surface RMS), a set of flattening currents was determined using an iterative process.    At each step, an image was taken using the Zygo interferometer. The next desired position for the actuators to reproduce in the iteration ($i$) was set to the negative of the last image taken ($S_i$).
The next set of currents ($F_i$) to produce the desired shape ($S_i$) were calculated using the influence functions matrix ($P$) through the  Moore-Penrose inverse: 

\begin{equation}
  F_i = (P^{T}P)^{-1} P^T S_i.
\end{equation}

\noindent The currents calculated for each iteration were summed with the previous currents applied to find the new values to be applied ($F = \sum_{1}^{i = n} F_i$). 

\begin{figure}[ht]
\centering
\begin{subfigure}{0.8\textwidth}
  \centering
  \includegraphics[width=0.85\linewidth]{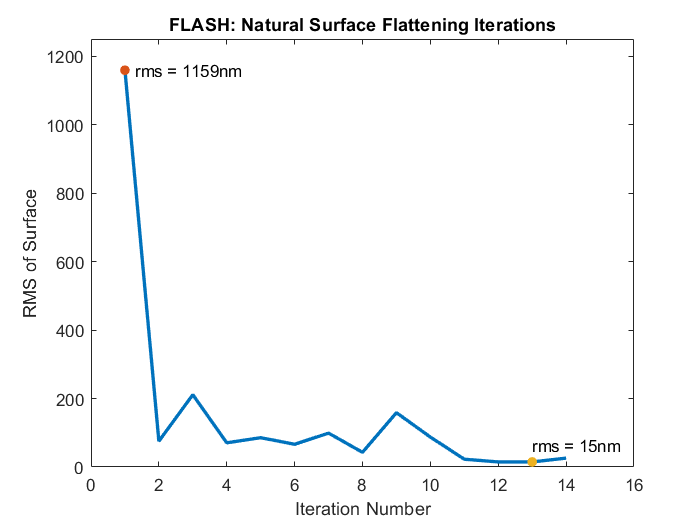}
\end{subfigure}
\begin{subfigure}{1.0\textwidth}
  \centering
   \includegraphics[width=1.0\linewidth]{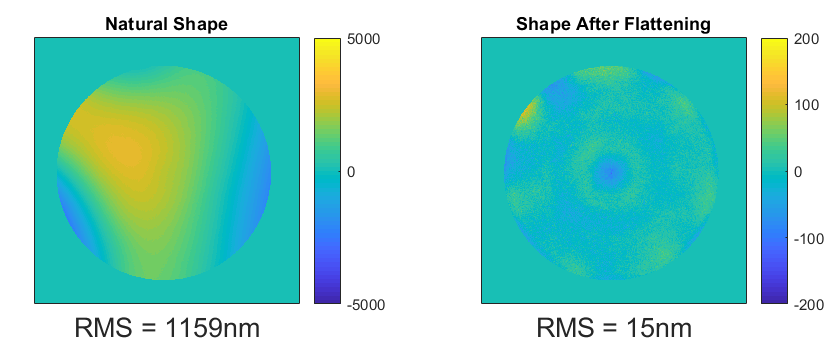}
 \end{subfigure}
 \caption[example]{\textbf{Natural-Shape Surface Flattening}. (\textit{Top}) \emph{RMS of the Mirror Surface Shape during Flattening Iterations}.  The best set of flattening currents was determined through an iterative process of 14 steps.     (\textit{Bottom left}) \emph{FLASH Natural Surface Shape}. Unpowered, the shape of the TNO FLASH mirror surface has a peak-to-valley that is 5339nm ($RMS = 1159$nm within \diameter140mm).  (\textit{Bottom right}) \emph{Surface Shape after Applied Flattening}.
  After flattening, the surface shape was brought down to 15nm RMS with a peak-to-valley of 261nm. The colorbar is reported in units of nanometers. }
   \label{fig:flattening}
   \end{figure}

 On April 5th 2021, fourteen iterations of flattening were completed to calculate the optimal set of flattening currents (Figure \ref{fig:flattening}, \textit{top}).  The majority of the variation was removed after four iterations. The flattest shape was found on iteration thirteen which had an RMS $= 14.7$nm (within $\diameter$140mm) with a peak-to-valley of 260.0nm. The surface RMS was reduced by a factor of 79.  
 Figure \ref{fig:flattening} (\textit{bottom}) shows the natural shape of the mirror alongside the shape after this flattening. 
 The total current needed to hold this shape across the 19 actuators was 323.9mA, averaging 17.0 $\pm$ 4.7mA per actuator. The required power was 7.3mW (0.4mW/actuator). This result confirms that the use of a passive air cooling will be appropriate for the UH2.2m adaptive secondary mirror. 
 

\subsection{Linearity} 

\emph{Measurement setup: Zygo Interferometer \& Capacitive Sensors}

\begin{figure}[h!]
 \centering
  {\includegraphics[width=0.5\textwidth]{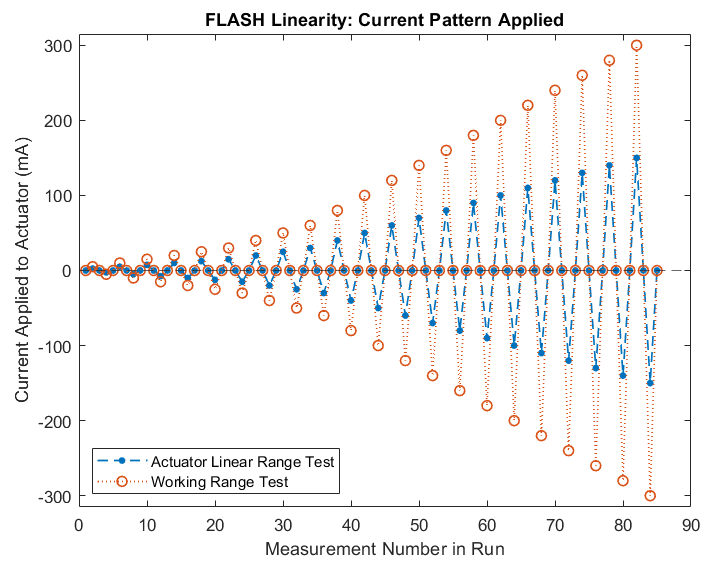}}~    
  {\includegraphics[width=0.45\textwidth]{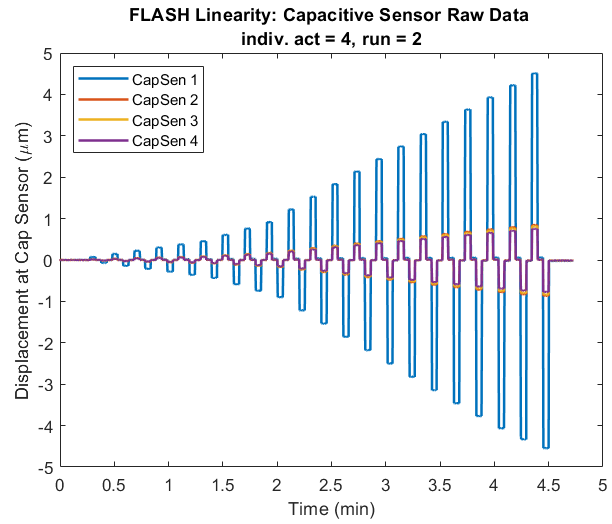}}~
   \caption { \label{fig:CurrentPatterns} 
\textbf{Linearity Testing} (\textit{Left}) The linearity testing was completed using currents varying across the actuators linear range ($\pm150$mA) and full working range ($\pm300$mA). (\textit{Right}) The four capacitive sensors were used to collect real time displacement data during the linearity testing.  This example data set was taken during an individual actuator run of Actuator 4.   }
   \end{figure}

 Linearity testing was performed by poking actuators individually (using Actuators 2, 4, 6, and 13) and as a group simultaneously (all actuators).  The individual actuator test  measured the high spacial order performance.  The all-actuator test demonstrated the capabilities of the HVR actuators if  minimal inter-actuator and shell deformation forces exist.    
 The actuators were run through two current patterns which applied positive and negative currents from $\pm150$mA and $\pm300$mA  (Figure \ref{fig:CurrentPatterns}). The first current pattern probed the HVR actuators quoted linear range, from 0 to $\pm$150mA. The second pattern spanned the maximum possible working range of the HVR actuators when controlled with the analog electronics, from 0 to $\pm$300mA.  
 
 Four runs were completed for each configuration.  
 Zygo interferometer data and capacitive sensor data were collected simultaneously during each run. The Zygo interferometer data was used to analyze the individual actuator tests because the capacitive sensors are not able to detect the height of the mirror surface at the actuator's center.  The capacitive sensors were used to analyze the all-actuator tests because the Zygo interferometer cannot measure a change in piston.

\begin{figure}[h!]
 \centering
  {\includegraphics[width=0.55\textwidth]{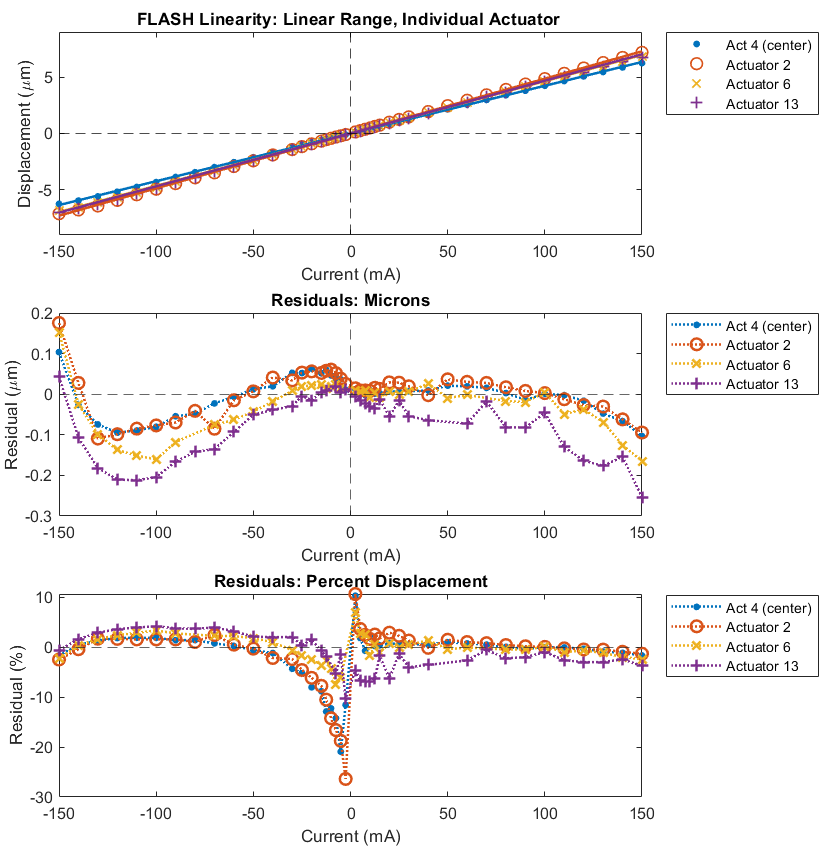}}~    
  {\includegraphics[width=0.45\textwidth]{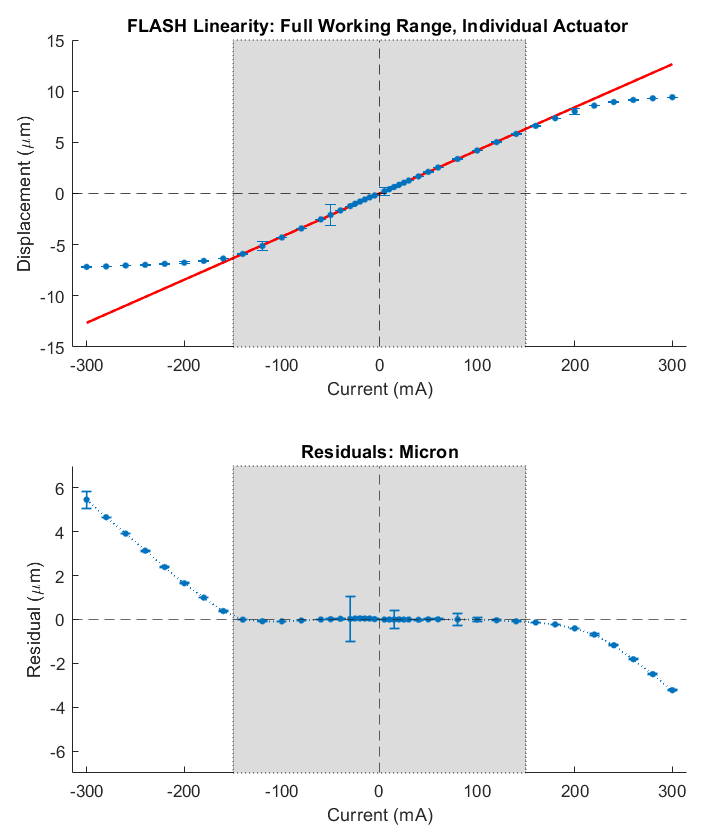}}~
   \caption { \label{fig:NonLinInd} 
\textbf{Linearity Measurements from the Zygo Interferometer for the Actuators Poked Individually}. (\textit{Left}) The linearity measured from the linear-range test ($\pm150$mA) for Actuators 2, 4, 6 and 13 was $lin =  95.2 \pm 0.8$\%.  The average displacement was $46.5 \pm 1.4 $nm/mA.   (\textit{Right}) The displacement measured by the Zygo interferometer data from five runs of Actuator 4 across its working range (blue) was averaged and fit with a zero-intercept linear fit (red).  The working range-test confirms that the linear region of the actuators is approximately $\pm150$mA (gray), but the displacements are asymmetric between the positive and negative applied currents. }
   \end{figure}

\begin{figure}[h!]
 \centering
  {\includegraphics[width=0.49\textwidth]{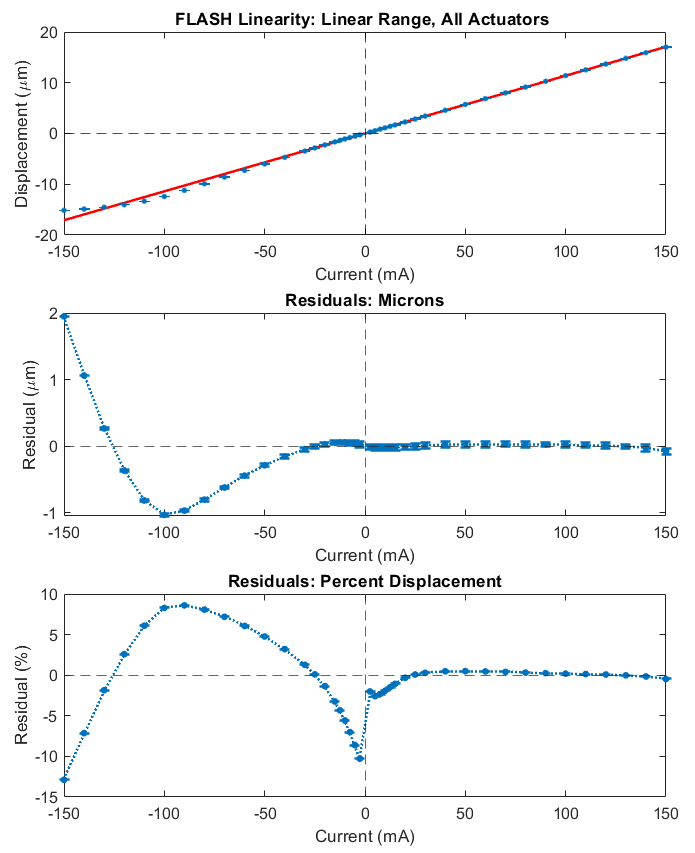}}~    
  {\includegraphics[width=0.49\textwidth]{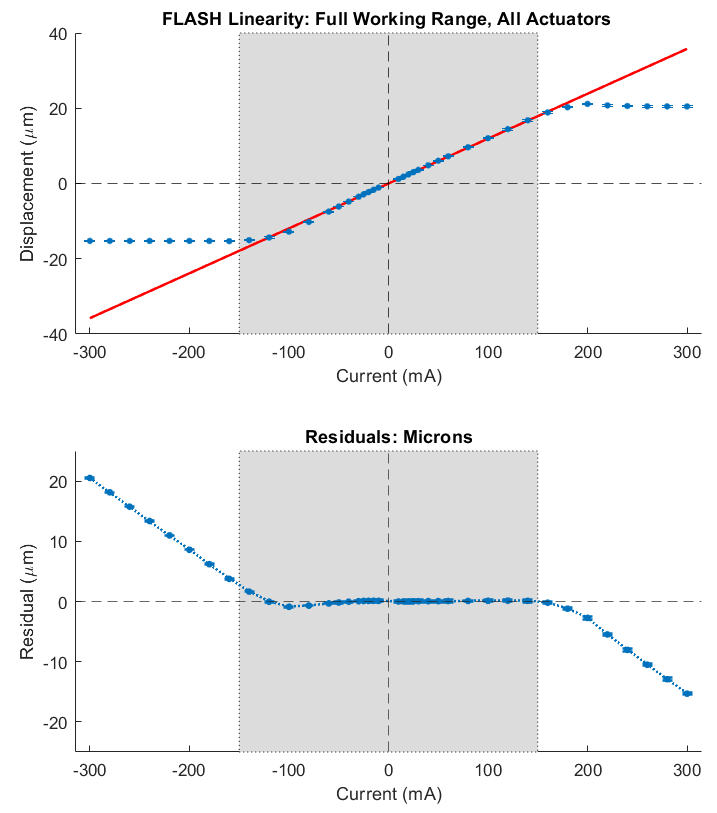}}~
   \caption { \label{fig:NonLinALL} 
\textbf{Linearity Measurements from the Capacitive Sensors when All Actuators were Moved Synchronously} (\textit{Left}) The displacement measured by the capacitive sensor data from three runs (blue) was averaged and fit with a zero-intercept linear fit (red).  The linearity measured from the linear-range testing ($\pm150$mA) when all actuators were moved simultaneously was $lin =  96.8 \pm 0.5$\%.  The average displacement was $113.9 \pm 1.0 $ nm/mA.   (\textit{Right}) The working-range test ($\pm300$mA) when all actuator were moved simultaneously demonstrates that the actuator range cannot extend beyond TNO's quoted linear range ($\pm150mA$), which is shaded in gray.  }
   \end{figure}

Table \ref{tab:nonlin} summarizes the key results from our linearity testing. Our testing reveals that within the stated linear range of the actuator, the high-order linearity is $95.2\pm 0.8$\% with a max residual of 0.2$\mu$m surface error (Figure \ref{fig:NonLinInd}).  The linearity measured when all-actuators were moved simultaneously was $96.8 \pm 0.5$\% with a max residual of 2$\mu$m (Figure \ref{fig:NonLinALL}).  The average displacement measured for the all-actuator movement was roughly 2.5 times larger than the average displacement measured for the individual actuator movements.  All testing showed an asymmetry between positive and negative applied currents, with larger movements available when positive currents are applied.

\begin{center}
\begin{table}[h!]
\centering
\caption{\textbf{FLASH Linearity Results Summary}}
\label{tab:nonlin}
\begin{tabular}{|c|c|c|c|c|c|}
\hline
\textbf{\begin{tabular}[c]{@{}c@{}}Individual act/\\ All actuators\end{tabular}} & \textbf{\begin{tabular}[c]{@{}c@{}}Current \\ Range\end{tabular}} & \textbf{\begin{tabular}[c]{@{}c@{}}Actuators \\ Used\end{tabular}} & \textbf{\begin{tabular}[c]{@{}c@{}}Average \\ Displacement\end{tabular}} & \textbf{\begin{tabular}[c]{@{}c@{}}Max \\ Displacement\end{tabular}}                        & \textbf{\begin{tabular}[c]{@{}c@{}}Percent \\ Linearity\end{tabular}} \\ \hline
\multirow{2}{*}{Individual}                                                      & \begin{tabular}[c]{@{}c@{}}Linear \\ ($\pm$150mA)\end{tabular}    & 2,4,6,13                                                           & $46.5 \pm 1.4 $ nm/mA                                                    & \begin{tabular}[c]{@{}c@{}}$-6.8\mu$m @ $\pm$-150mA\\  $6.8\mu$m @ $\pm$+150mA\end{tabular} & $95.2 \pm 0.8$\%                                                      \\ \cline{2-6} 
                                                                                 & \begin{tabular}[c]{@{}c@{}}Working \\ ($\pm$300mA)\end{tabular}   & 4                                                                  & -                                                                        & \begin{tabular}[c]{@{}c@{}}-7.2$\mu$m @ -300mA\\ 9.4$\mu$m @ +300mA\end{tabular}            & -                                                                     \\ \hline
\multirow{2}{*}{All}                                                             & \begin{tabular}[c]{@{}c@{}}Linear \\ ($\pm$150mA)\end{tabular}    & all                                                                & $113.9 \pm 1.0 $ nm/mA                                                   & \begin{tabular}[c]{@{}c@{}}-15.1$\mu$m @ -150mA\\ 17.0$\mu$m @ +150mA\end{tabular}          & $96.8 \pm 0.5$\%                                                      \\ \cline{2-6} 
                                                                                 & \begin{tabular}[c]{@{}c@{}}Working \\ ($\pm$300mA)\end{tabular}   & all                                                                & -                                                                        & \begin{tabular}[c]{@{}c@{}}-15.3$\mu$m @ -300mA\\ 20.5$\mu$m @ +300mA\end{tabular}          & -                                                                     \\ \hline
\end{tabular}
\end{table}
\end{center}

\subsection{Hysteresis}

\emph{Measurement setup: Zygo Interferometer \& Capacitive Sensors}

\begin{figure}[ht]
\centering
\begin{subfigure}{1.0\textwidth}
  \centering
  \includegraphics[width=1.0\linewidth]{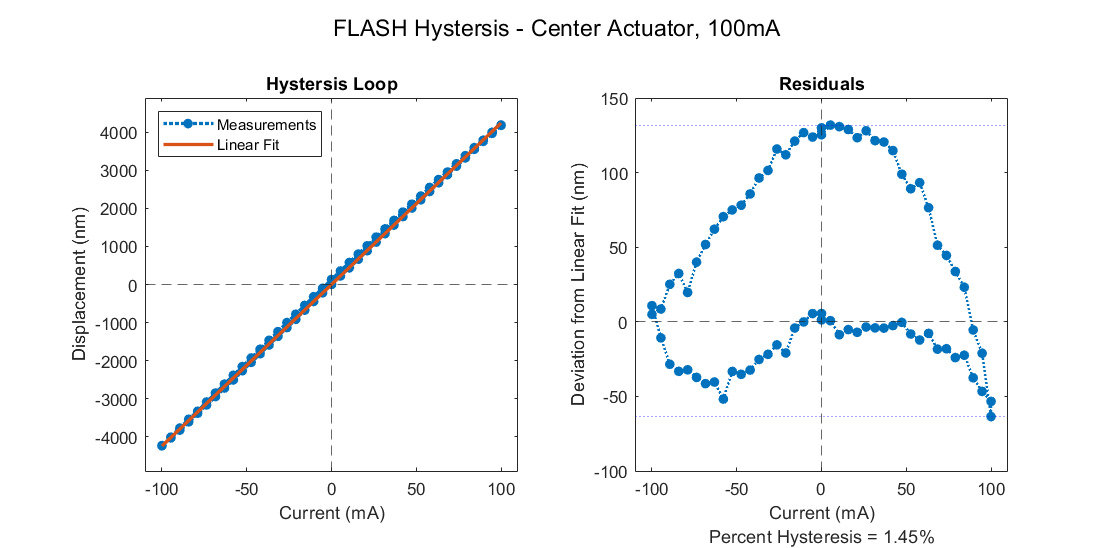}
  \label{fig:hystcurrentpattern}
\end{subfigure}
\begin{subfigure}{1.0\textwidth}
  \centering
   \includegraphics[width=0.75\linewidth]{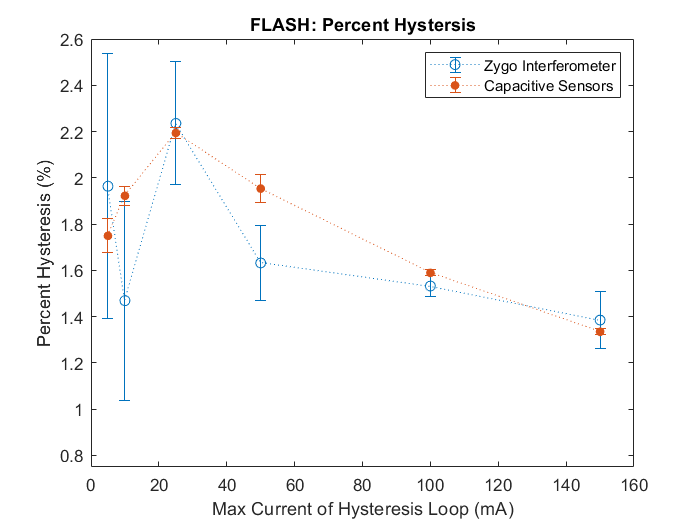}
 \end{subfigure}
 \caption[example]{\textbf{Hysteresis testing results}. (\textit{Top}) \emph{Example Hysteresis Loop} from Actuator 4 with max current of 100mA.   (\textit{Bottom}) \emph{Percent Hysteresis}. The data point at each current loop value are an average of the measurements from each the four actuators tested.  The average percent hysteresis was measured to be $1.80\pm0.13$\% using the Zygo data and $1.93\pm0.04$\% using the capacitive sensors within the definite linear range of the actuators (20mA - 120mA).     }
 \label{fig:hyst}
   \end{figure}

Actuator displacement varies depending on the previous position of the actuator.  To quantify this effect, hysteresis testing was performed using four actuators (2, 4, 6, and 13) poke individually, stepping through five independent hysteresis loops. Each loop was done in eighty steps between currents of $\pm$5mA, $\pm$10mA,  $\pm$25mA, $\pm$50mA, $\pm$100mA, $\pm$150mA.  Three runs were performed for each actuator at $\pm$5mA, $\pm$10mA,  $\pm$150mA and five runs were performed for each actuator at $\pm$25mA, $\pm$50mA, $\pm$100mA. An example of one of these hysteresis loops is shown in Figure \ref{fig:hyst}. 

We quantify the hysteresis using the definition: 

\begin{equation}
    Hyst(\%) = 100 \left(\frac{|S3-S1|}{|S2-S4|}\right)
\end{equation}

\noindent \noindent where $S1$ and $S3$ are the displacements measured at an applied current of zero and $S2$ and $S4$ are the displacements measured at the maximum and minimum current applied. 

Hysteresis data were collected simultaneously with the Zygo interferometer and capacitive sensors (Figure \ref{fig:hyst}). 
The average percent hysteresis was measured using the data from the $\pm$25mA, $\pm$50mA, $\pm$100mA runs to be $1.80\pm0.13$\% by the Zygo interferometer and $1.93\pm0.04$\% by the capacitive sensors. This calculation did not include the values measured outside the 20mA-120mA range to assure the hysteresis measurements were taken in the linear range of the actuators.  There was no dependency on the actuator tested and the measured hysteresis value. 

\subsection{Drift}

\emph{Measurement setup: Zygo Interferometer \& Capacitive Sensors}

Nine drift tests were performed for a duration of 8.5 hours with the goal of determining if the FLASH could hold its surface shape consistently across many hours of applied current. Table \ref{tab:drift} lists the type of current pattern applied for each drift test.  Zygo interferometer and capacitive sensor data were collected simultaneously.  The room temperature was also recorded using an Elitech RC-5 USB Temperature Data Logger during two runs. 

\begin{figure}[ht]
\centering
\begin{subfigure}{1.0\textwidth}
  \centering
  \includegraphics[width=0.75\linewidth]{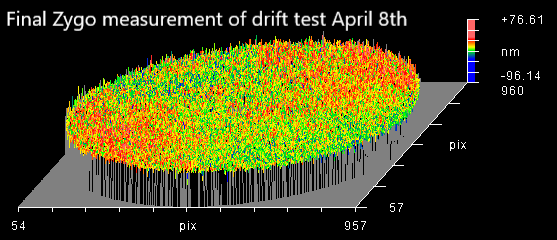}
\end{subfigure}
\begin{subfigure}{1.0\textwidth}
  \centering
   \includegraphics[width=0.75\linewidth]{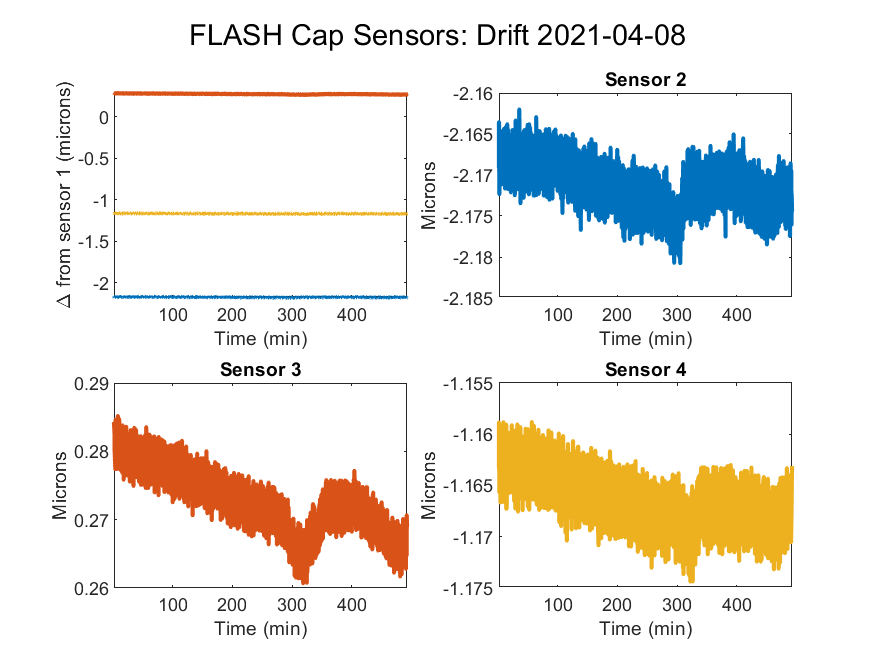}
 \end{subfigure}
 \caption[example]{\textbf{Example Drift Results from Holding the Flattening Currents Pattern}. (\textit{Top}) The final Zygo interferometer measurement could not measure  drifting outside the measurement error of the Zygo.   (\textit{Bottom}) Four capacitive sensors provided real time data over the course of the 8.5hr drift tests. The drifting due to a CTE mismatch of the capacitive sensor mounts was calibrated by subtracting the Capacitive Sensor 1 data from sensors 2-4.  In this run, the largest drift measured was $12$nm by Capacitive Sensor 3. }
 \label{fig:drift}
   \end{figure}

No drifting above the measurement error of the Zygo interferometer was found on the surface of the FLASH at the end of the test  (Figure \ref{fig:drift}, \textit{top}).  The capacitive sensors measured drifting of up to 90nm of change. However, these capacitive sensor drifts have a correlation with the recorded temperature data. These movements are in the range of what could be explained by the thermal expansion change in the 60mm aluminium capacitive sensor mounts.  To better calibrate out any movement due to temperature changes, the data from Capacitive Sensors 2-4 were subtracted from the data from Capacitive Sensor 1. After this correction, the biggest change recorded was 20nm of drift (Figure \ref{fig:drift}, \textit{bottom}).  Future sensor mount designs should consider using a material with a lower CTE value to increase the measurement precision of the capacitive sensors without this calibration.

\begin{table}[h!]
\centering
\caption{\textbf{Drift Testing Runs}}
\label{tab:drift}
\begin{tabular}{|c|c|c|l|}
\hline
\textbf{Drift Test} & \textbf{Pattern}                                                           & \textbf{Magnitude} & \multicolumn{1}{c|}{\textbf{Note}}                                                                                 \\ \hline
1                   & Checkerboard                                                               & +50mA             &                                                                                                                    \\ \hline
2                   & Checkerboard                                                               & +100mA             &                                                                                                                    \\ \hline
3                   & Piston                                                                     & +100mA             &                                                                                                                    \\ \hline
4                   & Piston                                                                     & -100mA             & \multicolumn{1}{c|}{}                                                                                              \\ \hline
5                   & Piston                                                                     & +50mA             & \multicolumn{1}{c|}{}                                                                                              \\ \hline
6                   & \begin{tabular}[c]{@{}c@{}}Flattening Currents \\ found 04-07\end{tabular} & --                 &                                                                                                                    \\ \hline
7                   & \begin{tabular}[c]{@{}c@{}}Flattening Currents\\  found 04-07\end{tabular} & --                 & \multicolumn{1}{c|}{\begin{tabular}[c]{@{}c@{}}Taken on 04-08-2021, \\ results shown in Figure \ref{fig:drift}\end{tabular}} \\ \hline
8                   & Checkerboard                                                               & -50mA             &                                                                                                                    \\ \hline
9                   & Checkerboard                                                               & -100mA             &                                                                                                                    \\ \hline
\end{tabular}
\end{table}

   \clearpage

\subsection{Zernike Mode Testing} 

\emph{Measurement setup: Zygo Interferometer \& Capacitive Sensors}

To determine how closely the FLASH can reproduce an applied Zernike shape,  the first 19 Zernike Modes were applied to FLASH.  The currents needed to reproduce the Zernike modes were calculated using Equation 2, where $S_i$ was the shape of each Zernike pattern. No iterations were used in trying to reproduce the Zernike pattern. 

\begin{figure}[ht]
   \begin{center}
   \begin{tabular}{c}
   \includegraphics[width=0.85\textwidth]{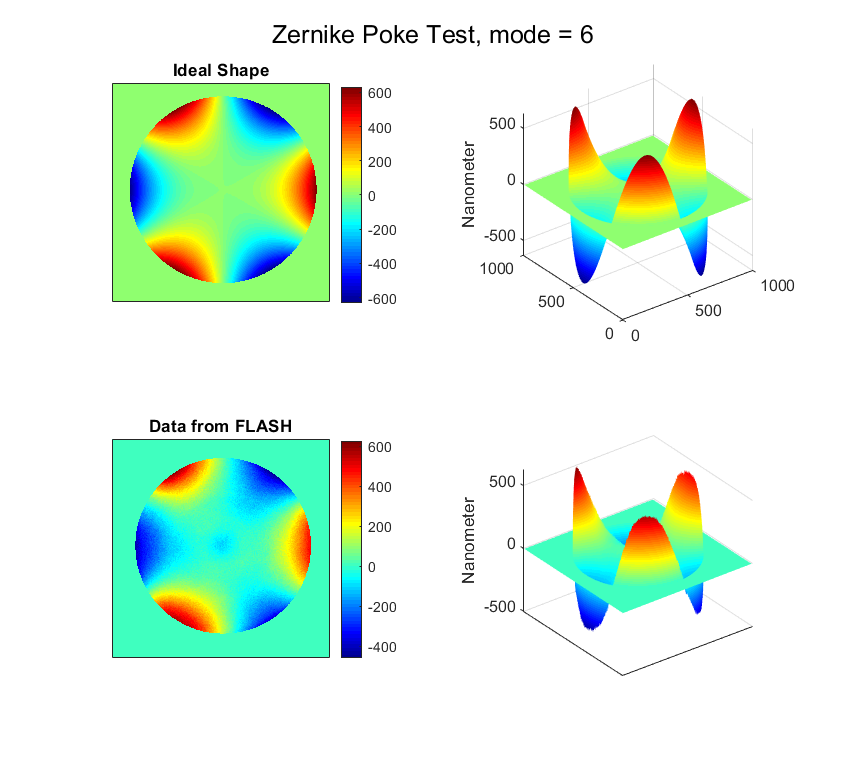}
   \end{tabular}
   \end{center}
   \caption[example] 
   { \label{fig:zernshape} 
   \textbf{Example Zernike Mode}: Zernike Mode 6 applied on FLASH. (\textit{Top}) The Zernike pattern fed to the FLASH. (\textit{Bottom}) The resulting facesheet pattern produced by FLASH. }
   \end{figure}

An example of the side-by-side comparison between DM3 and an ideal Zernike pattern can be viewed in Figure \ref{fig:zernshape} for mode 6.  A movie of the simultaneous capacitive sensor and Zygo interferometer data collection can be watched at the following youtube link: \href{https://youtu.be/scUhmqljJVc}{{\color{blue}https://youtu.be/scUhmqljJVc}}.

     \clearpage

\subsection{Settling Time}
\emph{Measurement setup: Capacitive Sensors}

To quantify the response time for an actuator to reach its final position after a current step is applied, we used the capacitive sensor system to measure the actuator settling time.  A current step of $+$10mA was applied to Actuator 4 (the center actuator).  Our definition of settling time was adopted from Rochette et al. 2018~\cite{Rochette2018}, which began/ended the settling time measurement when the system rose/reached 5\% of its final displacement. Ten runs were averaged to measure each settling time value.

We find a settling time of 14.3 $\pm$ 0.1 millisec when the current is applied as a step response (from 0 to 10mA in one step).
To overcome the actuators internal eddy-current behavior and speed up the settling time, a lead filter can be implemented when applying the current step (discussed here in Section 3.7) or the PI control settings can be tuned (discussed in Section 3.8). 

We implemented a lead filter using the discrete-space state Matlab Simulink model 
 with zero on the pole location of the measured plant and a pole at 500Hz. This filter applied the current using a spike and then a ramp-down from the peak.  It was designed to reach a peak current of 9.058 times the final current; for a step of 10mA, this resulted in a peak current of 90.58mA. 
 
\begin{figure}[ht]
   \begin{center}
   \begin{tabular}{c}
   \includegraphics[width=0.98\textwidth]{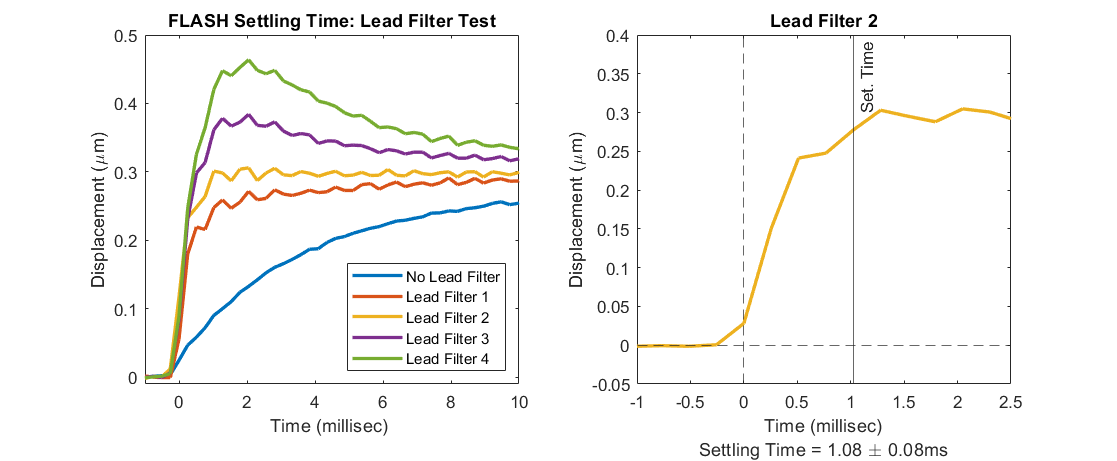}
   \end{tabular}
   \end{center}
   \caption[example] 
   { \label{fig:settlingtime} 
   \textbf{Settling Time} By applying a lead filter to the actuators, the response time of the actuator was improved by an order of magnitude from $t_{set} = 14.3 \pm 0.1$ms to  $t_{set} = 1.08 \pm 0.08$ms.  }
   \end{figure}

Four tuning values for the lead filter were tried with different decay times (Figure \ref{fig:settlingtime}).  The lead filter with the optimal tuning (Lead Filter 2) was set with the parameters of $A= 0.6$, $B= 2.18$, $C= -1.478$, $D = 9.058$.  The settling time measured with this lead filter was $t_{set} = 1.08 \pm 0.08$ ms, approximately an order of magnitude faster than the settling time with no lead filter applied.

\subsection{Dynamic Testing}

To verify the dynamic behavior of the FLASH DM, a non-parametric identification was performed by applying a random input signal to the center actuator and capturing the facesheet displacements.  This testing was conducted at TNO using a capacitive sensor system external to the FLASH (separate from the capacitive sensor system described here in Section 2.5).

\begin{figure}[ht]
   \begin{center}
   \begin{tabular}{c}
   \includegraphics[width=0.9\textwidth]{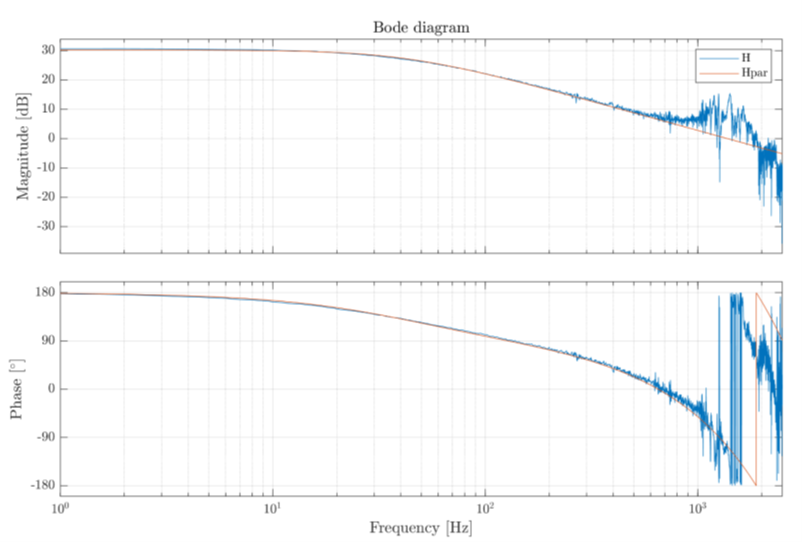}
   \end{tabular}
   \end{center}
   \caption[example] 
   { \label{fig:bode} 
   \textbf{Dynamical Test Bode Plot}  (\textit{Top}) The non-parametric identification of the transfer function (\textit{Bottom}) The first order parametric model fit to estimate pole location and IO delay. }
   \end{figure}

The frequency response of the transfer function measurement reveals a first order drop with a roll-off frequency of $\sim$40Hz  (Figure \ref{fig:bode}). The first second-order mechanical resonance is $\sim$1.2kHz. The observed first order lowpass response can be explained by the eddy-current behavior within the magnetic circuitry from the fast-changing currents in the coil. 
The open-loop transfer function shows the dynamical response of a first order parametric model that matches the first order low-passing behavior. A single pole at 42.6Hz and a two-sample delay at 5kHz are present, which is expected from the data acquisition system that was used. The close match between the response of this parametric model and the measurement data in phase and amplitude verify that the low-passing behavior can be characterized as first-order low-pass. 

\begin{figure}[ht]
 \centering
  {\includegraphics[width=0.49\textwidth]{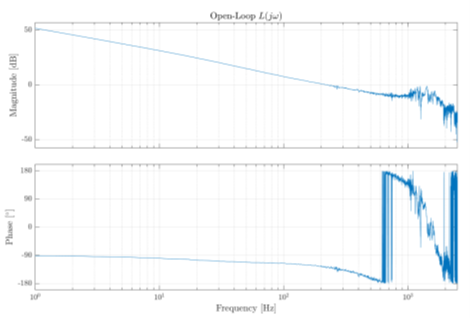}}~                
  {\includegraphics[width=0.49\textwidth]{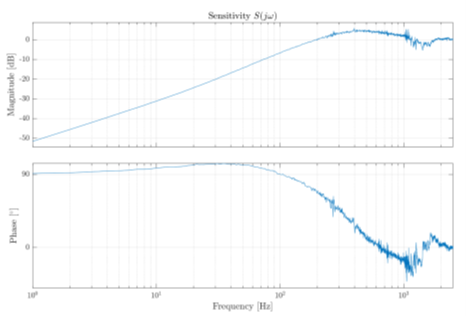}}~
   \caption { \label{fig:loopgain} 
\textbf{Loop-shaping Exercise} (\textit{Left}) Open-Loop Transfer function ($L(j\omega)$) (\textit{Right}) Sensitivity Transfer function ($S(j\omega)$).
   }
   \end{figure}

Although the first order low-pass behavior starts at 40Hz, the closed-loop control bandwidth can be  pushed to a faster frequency by properly tuning the PI controller. To demonstrate this, a loop-shaping exercise was performed based on the measured frequency responses using a PI controller. Figure \ref{fig:loopgain} shows the resulting Loop-gain ($L(j\omega)=H(j\omega)C(j\omega)$) and Sensitivity ($S(j\omega)=1/(1+L(j\omega))$) responses when tuning the PI controller to match the pole of the first order low-passing behavior within the actuator. With this PI controller, an open-loop bandwidth of 225Hz and a closed-loop sensitivity bandwidth of around 150Hz (-3dB) is achieved. Higher closed-loop performances are deemed possible when high-order controller design methods are employed. 

\clearpage
\section{CONCLUSIONS}

The key results from the UCSC-LAO initial performance testing of the FLASH are summarized in Table 4.  Our linearity, hysteresis, drift, and Zernike mode testing verify that the performance of the  2020 generation of TNO HVR actuators performed as expected and suggest that the performance of the UH2.2m adaptive secondary mirror will behave as TNO's modeling predicts. 

The dynamic and settling time measurements provide the first measured results of how the TNO systems perform at kilohertz  speeds.  
 Without tuning the profile of the current applied, the TNO deformable mirrors have a settling time of $t_{set} = 14.3 \pm 0.1$ms.  Our testing demonstrates that the correction speed can be improved to $t_{set} = 1.08 \pm 0.08$ms using a lead filter. This speed is consistent with the operating speed of the current on-sky adaptive secondary mirrors constructed with voice-coil style actuators.~\cite{Biasi2010}

The Zygo interferometer and capacitive sensor data were consistent when simultaneous data collection was performed. The two data collection methods complemented each other; the Zygo interferometer provided a holistic view of the mirror shape and the capacitive sensors provided dynamic time response information. The capacitive sensors are capable of measuring displacements at a higher precision than the Zygo interferometer, as long as having limited spatial information is acceptable.  In future situations when an interferometer is not possible to use in the testing setup (i.e. in an environmental chamber or on-sky optical system), our testing indicates that the capacitive sensors can accurately provide feedback about the real time performance of the deformable mirror.

\begin{table}[h!]
\centering
\caption{\textbf{Summary of FLASH Testing Results} \label{tab:conclusions}
}
\begin{tabular}{|c|c|l|}
\hline
\textbf{Section} & \textbf{Test}                                                                            & \multicolumn{1}{c|}{\textbf{Takeaway}}                                                                                                                                                                                                                              \\ \hline
3.1              & \begin{tabular}[c]{@{}c@{}}Influence Functions \&\\ Actuator Cross-Coupling\end{tabular} & \begin{tabular}[c]{@{}l@{}}\textit{Measured actuator spacing} $= 39.4 \pm 0.4$mm,\\\textit{Actuator cross-coupling}  $= 34.2  \pm 1.0$\%\end{tabular}                                                                                                                                \\ \hline
3.2              & \begin{tabular}[c]{@{}c@{}}Natural Shape \\ Surface Flattening\end{tabular}              & \begin{tabular}[c]{@{}l@{}}The surface was brought from \textit{RMS} = 1158nm to \textit{RMS} = 15nm \\ with an \textit{average current per actuator} = $17.0 \pm 4.7$mA and \\ \textit{total power} $= 7.3$mW ($0.4$mW/actuator)\end{tabular}                                                              \\ \hline
3.3              & Linearity                                                                                & \begin{tabular}[c]{@{}l@{}}\textit{Individual act}: avg disp. $= 46.5\pm1.4$nm/mA , lin $=  95.2\pm0.8$\%\\ \textit{All actuators}: avg disp. $= 113.9 \pm 1.0 $ nm/mA, lin $=  96.8 \pm 0.5$\%\\ There are asymmetries between negative and positive applied currents \end{tabular} \\ \hline
3.4              & Hysteresis                                                                               & \begin{tabular}[c]{@{}l@{}}Average \textit{percent hysteresis} from loops with max current 20mA-120mA: \\ $1.80\pm0.13$\% (zygo), $1.93\pm0.04$\% (capacitive sensors)\end{tabular}                                                                                          \\ \hline
3.5              & Drift                                                                                    & \begin{tabular}[c]{@{}l@{}}Max drift seen from cap sensors in 9 room-temperature tests: \\ \textit{Drift} $< 20$nm\end{tabular}                                                                                                                                              \\ \hline
3.6              & Zernike Mode                                                                             & The applied zernike modes are consistent with the measured modes                                                                                                                                                                                                    \\ \hline
3.7              & Settling Time                                                                            & \begin{tabular}[c]{@{}l@{}} $t_{set}=14.3\pm0.1$ms;\\ A lead filter can be implemented to lower the actuator response time;\\ a well-tuned lead filter achieved $t_{set}=1.08\pm0.08$ms for a \\ step of 10mA ($\sim3\mu$m).\end{tabular}                                                     \\ \hline
3.8              & Dynamic Testing                                                                          & \begin{tabular}[c]{@{}l@{}}\textit{First-order roll-off frequency} $= 40$Hz; \\ \textit{Second-order mechanical resonance} = $1.2$kHz\end{tabular}                                                                                                                                    \\ \hline
\end{tabular}
\end{table}

\acknowledgments   

The authors would like to thank Max Baeten, Arjo Bos, Mark Chun, Wouter Jonker, Matthew Maniscalco, Joost Peters and the memebers of the Lab for Adaptive Optics at UC Santa Cruz  for their guidance and assistance in the completion of this work.

\bibliographystyle{spiebib}   
\bibliography{report}   

\end{document}